\documentclass[eqsecnum,aps,superscriptaddress,12pt ]{revtex4}
\usepackage{lscape}
\usepackage{graphicx}
\usepackage{rotating}

\begin{document}
\title{ {\it Ab initio} calculations of oscillator strengths
and lifetimes of low-lying states in Mo VI}
\author{Narendra Nath Dutta$^1$, Gopal Dixit$^1$, B. K. Sahoo$^2$ and  Sonjoy Majumder$^1$ \\
\vspace*{0.3cm}
{\it $^1$Department of Physics and Meteorology, Indian Institute of Technology-Kharagpur, Kharagpur-721302, India \\
$^2$ Theoretical Physics Divison, Physical Research Laboratory, Ahmedabad-380009, India\\} }

\date{\today}

\begin{abstract}
\noindent Relativistic coupled-cluster (RCC) calculations
have been performed to estimate the electromagnetic forbidden transition
probabilities, oscillator strengths and lifetimes of many low-lying states of
five times ionized molybdenum (Mo VI). Contributions from the Breit
interaction up to the first order of perturbation have been examined.
Our results are in good agreement with the available other reported theoretical
and experimental results. A long lifetime about $4.9854 \ s$ of the
first excited state, 4d $^{2}D_{5/2}$, has been predicted which can be a
very useful criteria in the doping process of thin films. Correlations trends
from various RCC terms to the transition amplitude calculations are discussed.
\end{abstract}
\maketitle

\section{Introduction}

Electromagnetic forbidden, both magnetic dipole (M1)
and electric quadrupole (E2), transitions of Mo VI are important for
temperature and density estimations of tokamak plasmas
\cite{feldman, herter}, especially in the collision-radiative
model \cite{fournier, quinet}. Long lifetimes of metastable states
are dominated by these forbidden transitions and these states are
generally difficult to observe in the laboratory plasmas due to strong
collisions. However, these forbidden transitions of Mo VI have been
observed in laboratory in electron spin resonance experiment
\cite{baur} and therefore, they must be one of the sources of
density estimations in astrophysical plasmas where collisions are
very low due to high dilute interstellar medium \cite{charro}. Accurate
estimation of abundances of molybdenum in the atmosphere of the
evolved stars are important to understand the stellar nucleosynthesis
\cite{Orlov}.

Hexavelent molybdenum, isoelectronic to rubidium with 4p$^6$4d as
ground state configuration, is generated by electron impact in the
atomic collision process. The electron-impact ionization of
multiply charged Mo ions, relevant to astrophysics and laboratory
plasma research, have also been investigated \cite{hathiramani}.
Recently, Fisker et al. have given the possibility of the origin
of the lightest isotope of molybdenum in proton rich type II
supernova \cite{fisker}. The necessity of accurate estimation of allowed
dipole transition strengths to find out their mixing these effects in the
dipole forbidden transitions in Mo VI is explicitly discussed by T.
Yamamoto \cite{yamamoto}. Again, the transition strength between the
fine structure states of 4d lavel can reflect the electronic structure
of Mo VI in crystal \cite{szotek}.

A few calculations have been carried out to study the on electric dipole
(E1) transitions in Mo VI over the last few decades using the mean-field
theory \cite{migdalek,zilitis}. More recently, J. Reader \cite{reader} has
estimated the E1 transition probabilities among low-lying states by
estimating transition strengths in the semiempirical approach with
the experimental excitation energies.

For this single reference system, Mo VI, we have performed
relativistic coupled-cluster (RCC) calculation with single (S),
double (D) and partial triple (T) excitations in the framework of
Fock space multi-reference (FSMR). Both the excitation energies and
transition probabilities are determined using this RCC method using
which lifetimes of many low-lying states are estimated.

\section{Theory and Method of Calculations}

\subsection{Theory}

The oscillator strength for E1 transition from $|\Psi_f\rangle$ to
$|\Psi_i\rangle$ is given as
\begin{equation}
f_{fi} ={2\over {3g_f}}\Delta E_{fi}\times |D_{fi}|^2 ,
\end{equation}\label{eq7}
where $\Delta E_{fi}$ is the excitation energy between the upper
and lower states and $g_f=2J_f+1$ is the degeneracy factor of the
upper state with total angular momentum $J_f$.

The single particle reduced matrix elements for the E1, E2 and M1
transition operators are given in \citep{johnson}. The emission
transition probabilities (in sec$^{-1}$) for the E1, E2 and M1
channels from states {\it f} to {\it i} can be expressed as
\begin{equation}
A^{E1}_{fi} =
\frac{2.0261\times10^{18}}{\lambda^{3}(2j_f+1)}S^{E1},
\end{equation}
\begin{equation} A^{E2}_{fi} =
\frac{1.11995\times10^{18}}{\lambda^{5}(2j_f+1)}S^{E2},
\end{equation}
\begin{equation}
A^{M1}_{fi} =
\frac{2.69735\times10^{13}}{\lambda^{3}(2j_f+1)}S^{M1},
\end{equation}
where $S^O = {|{\langle \Psi_f|O|\Psi_i\rangle}|}^2$ is the transition strength
for the coressponding operator $O$ (in a.u.) and $\lambda$ (in
\AA ) is the corresponding transition wavelength.

The lifetime of a particular excited state $i$ can be computed by the
reciprocal of the total transition probability, $\sum_{j} A_{ij}$ (in
sec$^{-1}$), arising from all possible states $j$ due to spontaneous
electromagnetic transitions, i.e.
\begin{equation}\label{eq513}
\tau_{i} = \frac {1}{\sum_{j}{A_{ij}}}.
\end{equation}

\subsection{Fock Space Multi-reference RCC theory}

The FSMRCC method is one of the most powerful highly correlated
many-body approaches due to its all order structure to account the
correlation effects \cite{lindgren}. The FSMRCC, which is mainly meant for
multi-reference systems, is used here for the one valence electron and
has been described in details elsewhere \cite{lindgren, mukherjee,
Haque, Pal}. Here we present the method briefly.

We first consider the Dirac-Coulomb Hamiltonian for a closed-shell
$N$ electron system which is given by
\begin{equation}\label{eq2.1}
{\mbox
H}=\sum_{i=1}^{N}\left[c\vec{\alpha_{i}}\cdot\vec{p}_{i}+\beta
mc^{2}
          +V_{\mathrm{Nuc}}(r_{i})\right]+\sum_{i<j}^{N}\frac{1}{r_{ij}}
\end{equation}
with all the standard notations often used.

The theory for a single valence system is based on the concept of
common vacuum for both the closed shell $N$- and open shell
$N\pm1$-electron systems, which allows us to formulate a direct
method to determine energy differences (electron attachment energy
or negative of the ionization potential). Also, the holes and particles are
defined with respect to the common vacuum for both the electron
systems. Model space of an (n,m) Fock-space contains determinants
of $n$ holes and $m$ particles distributed within a set of
orbitals known as {\em active} orbitals. For example, in the
present article, we are dealing with (0,1) Fock-space which is a
complete model space (CMS) by construction and is given by

\begin{equation}\label{eq2.4}
|\Psi^{(0,1)}_\mu\rangle=\sum_i {\mbox C}_{i\mu}
|\Phi_i^{(0,1)}\rangle,
\end{equation}
\noindent where ${\mbox C}_{i\mu}$'s are the expansion
coefficients of $\Psi^{(0,1)}_\mu$, and $\Phi^{(0,1)}_i$'s are the
model space configurations made of DF orbitals. The dynamical
electron correlation effects are introduced through the {\em
valence-universal} wave-operator $\Omega$ \cite{lindgren,
mukherjee}
\begin{equation}\label{eq2.5}
\Omega={\{ \exp({\tilde{S}}) }\},
\end{equation}
\noindent where
\begin{equation}\label{eq2.6}
{\tilde{S}}=\sum_{k=0}^m\sum_{l=0}^n{S}^{(k,l)}
                 ={S}^{(0,0)}+{S}^{(0,1)}+ {S}^{(1,0)}+\cdots
\end{equation}
At this juncture, it is convenient to single out the core-cluster
amplitudes $S^{(0,0)}$ and call them $T$. The rest of the cluster
amplitudes will henceforth be called $S$. Since $\Omega$ is in
normal ordered, we can rewrite Eq. (\ref{eq2.5}) as
\begin{equation}\label{eq2.7}
\Omega ={\mbox{exp}({T}) }{\{\mbox{exp}({S}) }\}.
\end{equation}
The ``valence-universal'' wave-operator $\Omega$ in Eq.
(\ref{eq2.7}) is parameterized in such a way that the states
generated by its action on the reference space satisfy the
Fock-space Bloch equation
\begin{equation}\label{eq2.8}
{\mbox H}\Omega {\mbox P}^{\mbox{(k,l)}}=\Omega {\mbox
P}^{\mbox{(k,l)}}
                    {\mbox H}_{\tiny\mbox {eff}} {\mbox
                    P}^{\mbox{(k,l)}},
\end{equation}
where
\begin{equation}\label{eq2.9}
{\mbox H}_{\tiny\mbox {eff}}={\mbox P}^{\mbox{(k,l)}}{\mbox H}
 \Omega {\mbox P}^{\mbox{(k,l)}}.
\end{equation}
Here, P is the projection operator of model space. Eq.
(\ref{eq2.8}) is valid for all (k,l) starting from k=l=0 (i.e.,
the {\em core} problem) to some desired {\em parent} model space,
with k=m, l=n. In this present calculation, we truncate Eq.
(\ref{eq2.6}) at $m=0$ and $n=1$.

In this work, single ($T_{1}, S_{1}^{(0,1)}$) and double ($T_{2},
S_{2}^{(0,1)}$) excitations are considered for $T$ and $S$
clusters operator. Therefore, the total correlated wavefunction of
the system with single valence orbital {\it v}, can be written as
\begin{equation}
|\Psi_v\rangle = \Omega |\Psi^{(0,0)}\rangle =
e^{T_{1}+T_{2}}\{1+S_{1v}^{(0,1)}+S_{2v}^{(0,1)}\}|\Psi^{(0,0)}\rangle.
\end{equation}
Important triple excitations, correspond to the correlation to the
valence orbitals, are included in the open shell FSMRCC-SD
calculations by an approximation that is similar in spirit to
FSMRCC-SD(T) \cite{ccsd(t)}. The approximate valence triple
excitation amplitudes are given by
\begin{equation}
{S_v}_{(0,1)abk}^{pqr}=\frac{{\{{\overbrace{V{T}_2}}\}_{abk}^{pqr}}+{\{{\overbrace{V{S_{2v}^{(0,1)}}}}\}_{abk}^{pqr}}}{\varepsilon_{a}+\varepsilon_{b}+\varepsilon_{k}-\varepsilon_{p}-\varepsilon_{q}-\varepsilon_{r}},
\end{equation}
where ${S_{v}}_{(0,1)abk}^{pqr}$ are the amplitudes corresponding to
the simultaneous excitations from core orbitals $a,b$ and valence
$k$ to virtual orbitals $p,q$, and $r$, respectively. $\overbrace{V{T}_2}$ and
$\overbrace{V{\mbox S_{2v}^{(0,1)}}}$ are the connected composites
involving $V$ and $T$, and $V$ and $S_{v}^{(0,1)}$, respectively,
where $V$ is the two electron Coulomb ($\frac{1}{r_{ij}}$)integral and
$\varepsilon$'s are the orbital energies.

The transition matrix element due to any operator $O$ can be
expressed as
\begin{eqnarray}
&&O_{fi} =  \frac{\langle \Psi_f|O|\Psi_i\rangle} {\sqrt{{\langle
\Psi_f|\Psi_f\rangle}
{\langle \Psi_i|\Psi_i\rangle}}} \nonumber \\
&=& \frac{{\langle \Phi_f|\{1+{S_f}^{\dag
{(0,1)}}\}{e^T}^{\dag}Oe^T\{1+S_i^{(0,1)}\}|\Phi_i\rangle}}
{\sqrt{{\langle \Phi_f|\{1+{S_f}^{\dag
{(0,1)}}\}{e^T}^{\dag}e^T\{1+S_f^{(0,1)}\}|\Phi_f\rangle} {\langle
\Phi_i|\{1+{S_i}^{\dag
{(0,1)}}\}{e^T}^{\dag}e^T\{1+S_i^{(0,1)}\}|\Phi_i\rangle}}}. \nonumber \\
\end{eqnarray}

\section{Results and Discussions}
\begin{figure}
\begin{center}
\includegraphics[width=0.8\textwidth]{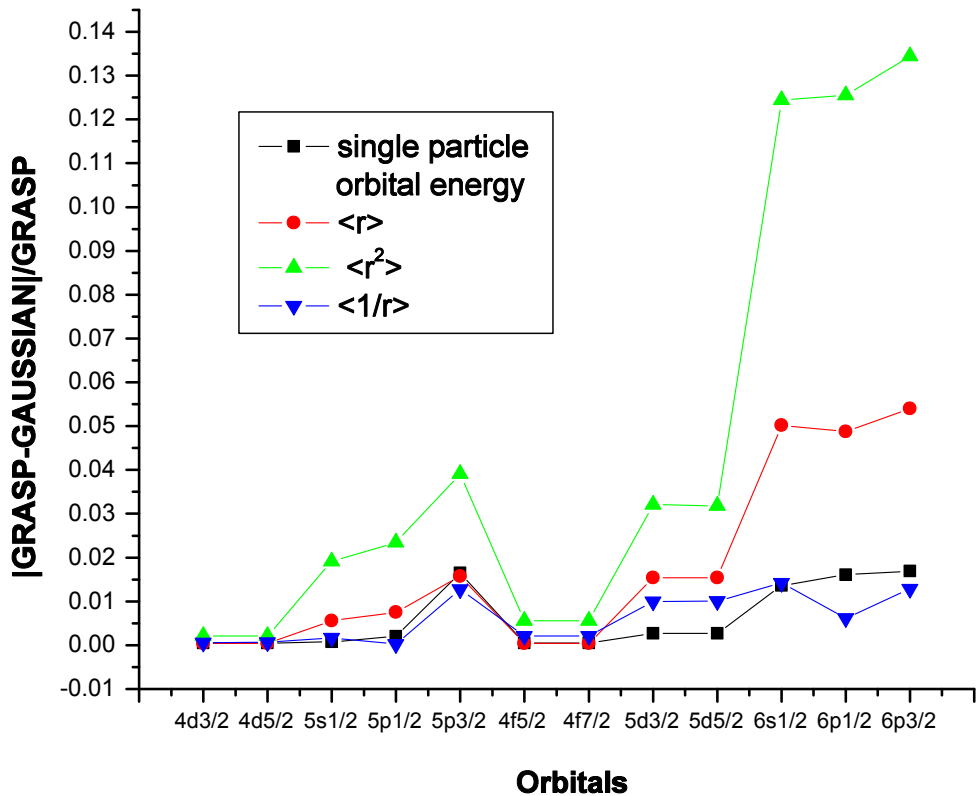}
\caption{\label{label}The relative energies and expectation values
of $r$, $r^2$ and $1/r$ of the DF GTO orbitals to the DF GRASP
orbitals.} \label{fig1}
\end{center}
\end{figure}

In the present calculation,  the radial wavefunctions of DF
orbitals of closed shell Mo VII are obtained using Gaussian type
orbitals (GTO) basis with finite nuclear size as discussed in
our earlier paper \cite{gopal}. We have used universal basis
set, where the exponent $\alpha_{i}$ is related with two
parameters $\alpha_{0}$ and $\beta$, same for all the
symmetries, expressed as
\begin{equation}
\alpha_{i}=\alpha_{0}\beta^{i-1}.
\end{equation}
We have considered $\alpha_{0}$ and $\beta$ as 0.00625 and 2.72,
respectively, after obtaining best fit of the bound orbital
energies and evaluating the expectation values of different radial
functions ($r$ , $r^2$ , $1/r$ ) generated with GTOs and GRASP2
\cite{parpia}. In the DF calculations, we have taken 22, 20, 17,
15, and 12 number of GTOs for s, p, d, f, and g type symmetries,
respectively, to generate the atomic orbitals. In Fig. \ref{fig1},
we have given the relative errors obtained for different orbitals
in the calculations of these quantities using the above chosen
parameters. Since these errors are very small, it shows that there
is a good agreement between results. We assume that both the bound
and continuum orbitals generated using the above parameters will
describe well both inside and outside of the nucleus. Therefore,
we have considered all the orbitals obtained using GTOs for the
rest of the calculations.

The number of the DF orbitals for different symmetries used in
the present calculation is based on convergent criteria of core
correlation energy of Mo VII for which it satisfies numerical
completeness. The number of DF orbitals considered for s, p, d, f, and
g type symmetries in the RCC calculations are 12, 11, 10, 9 and 8,
respectively; and among them 9, 8, 7, 5, and 5 are bound orbitals,
respectively, including all the core orbitals. The $T$ amplitudes are
first determined by solving the closed shell RCC equations for the
closed-shell system (Mo VII), then $S$ amplitudes are solved
from the open-shell equations for the single-valence states of Mo
VI.

Table I summarizes the calculated excitation energies (EE) and
fine structure splitting (FS) of low-lying excited states and
their comparison with the recent experimental results \cite{reader}.
The average deviation is around 0.5\% for EE. We have also
presented the contribution from the partial triple excitations to
the EE (E$_{\textrm{triple}}$), which is around 0.3\% to the total
EE. We have examined the first order excitation energy corrections
due to Breit interaction using large scale relativistic CI
calculations. Maximum contribution is coming for 5s state, which
is around $+2\%$, whereas, contributions to 5p and 4f states  are
around $+0.4\%$ and $-0.04\%$; respectively. For 5d state, it is
as small as 54 $cm^{-1}$ consistent with the result obtained by Pan
and Beck \cite{pan}. Since, the contributions due to Breit
interaction are relatively small, we do not consider them here
selfconsistently to evaluate wavefunctions.

\begin{table}
\caption{Excitation energies (EE) and fine structure splitting
(FS) (in cm$^{-1}$) of Mo VI and their comparison with
experimental results. Contribution from  absolute value of partial
triple excitation to the EE (E$_{\textrm{triple}}$) are also
presented.}
\begin{tabular}{lrrrrr}
\hline
      & \multicolumn {2}{c} {EE} & \multicolumn {2}{c} {FS} &  EE$_{\textrm{triple}}$ \\
State & CC  & Exp.$^a$ & CC  & Exp.$^a$ &   \\ \hline \hline
4d $^{2}D_{3/2}$ & 0 &  0 & & & 0\\
4d $^{2}D_{5/2}$ & 2670.55 & 2583.50 & 2670.55 & 2583.50 & 572.36 \\
5s $^{2}S_{1/2}$ & 118536.63 & 119725.62 & & & 396.72 \\
5p $^{2}P_{1/2}$ & 181795.01 & 182404.47 & & & 220.12 \\
5p $^{2}P_{3/2}$ & 186825.31 & 187331.19 & 5030.30 & 4926.72 & 235.92 \\
4f $^{2}F_{5/2}$ & 269411.28 & 267047.22 & & & 492.50\\
4f $^{2}F_{7/2}$ & 269653.95 & 267456.84 & 242.67 & 409.62 & 621.98 \\
5d $^{2}D_{3/2}$ & 282598.52 & 282825.59 & & & 697.69\\
5d $^{2}D_{5/2}$ & 283396.42 & 283610.94 & 797.90 & 785.35 & 816.42 \\
6s $^{2}S_{1/2}$ & 315632.41 & 313806.81 &        &        & 857.47 \\
6p $^{2}P_{1/2}$ & 342531.49 & 340570.78 &        &        & 690.67\\
6p $^{2}P_{3/2}$ & 344853.69 & 342562.44 & 2322.21& 1991.66& 697.91 \\
 \hline
\end{tabular}
\label{tab:results1}

$^a$ Ref. \cite{reader}. \\
\end{table}

Since the transition rate is proportional to the square of the
transition amplitude, therefore precise description of the
wavefunction is necessary due to one order higher dependance on
wavefunctions than energy. In Table II, we compare the E1
transition amplitude in both length and velocity gauges for few transitions. We find a
good agreement between them, which is one of the
characteristics to judge the accuracy of the wavefunctions.

\begin{table}
\caption{Absolute values of E1 transition amplitude in length
(D$_l$) and velocity (D$_v$) gauges for Mo VI.}
\begin{tabular}{lcrrrr}
\hline
Term &    &    &             &        &        \\
Upper & & Lower  & D$_l$  & D$_v$  \\ \hline \hline
5p $^{2}P_{1/2}$ & $\rightarrow$ & 4d $^{2}D_{3/2}$ & 0.9851 & 0.8341 \\
5p $^{2}P_{1/2}$ & $\rightarrow$ & 5s $^{2}S_{1/2}$ & 1.7554 & 1.7522 \\
5p $^{2}P_{3/2}$ & $\rightarrow$ & 4d $^{2}D_{3/2}$ & 0.4288 & 0.3619 \\
5p $^{2}P_{3/2}$ & $\rightarrow$ & 5s $^{2}S_{1/2}$ & 2.4885 & 2.5678 \\
4f $^{2}F_{5/2}$ & $\rightarrow$ & 4d $^{2}D_{3/2}$ & 1.3928 & 1.5438 \\
4f $^{2}F_{5/2}$ & $\rightarrow$ & 4d $^{2}D_{5/2}$ & 0.4272 & 0.4474 \\
4f $^{2}F_{7/2}$ & $\rightarrow$ & 4d $^{2}D_{5/2}$ & 2.6555 & 2.5330 \\
5d $^{2}D_{3/2}$ & $\rightarrow$ & 5p $^{2}P_{1/2}$ & 2.7486 & 2.6942 \\
5d $^{2}D_{3/2}$ & $\rightarrow$ & 5p $^{2}P_{3/2}$ & 1.2593 & 1.2278 \\
5d $^{2}D_{5/2}$ & $\rightarrow$ & 5p $^{2}P_{3/2}$ & 3.9600 & 3.7900 \\
6s $^{2}S_{1/2}$ & $\rightarrow$ & 5p $^{2}P_{1/2}$ & 0.9771 & 0.8873 \\
6s $^{2}S_{1/2}$ & $\rightarrow$ & 5p $^{2}P_{3/2}$ & 1.4711 & 1.1963 \\
6p $^{2}P_{1/2}$ & $\rightarrow$ & 4d $^{2}D_{3/2}$ & 0.1984 & 0.1686 \\
6p $^{2}P_{1/2}$ & $\rightarrow$ & 5d $^{2}D_{3/2}$ & 2.5559 & 2.2828 \\
6p $^{2}P_{1/2}$ & $\rightarrow$ & 6s $^{2}S_{1/2}$ & 3.4776 & 3.1844 \\
6p $^{2}P_{3/2}$ & $\rightarrow$ & 4d $^{2}D_{5/2}$ & 0.2845 & 0.2605 \\
6p $^{2}P_{3/2}$ & $\rightarrow$ & 5d $^{2}D_{3/2}$ & 1.1055 & 0.9949 \\
6p $^{2}P_{3/2}$ & $\rightarrow$ & 6s $^{2}S_{1/2}$ & 4.9149 & 4.6861 \\
\hline
\end{tabular}
\label{tab:results1}
\end{table}

In Table III, we compare our {\it ab initio} oscillator strength
values correspond to E1 transitions with the recent semi-empirical
calculations by Pan and Beck \cite{pan} and by Reader
\cite{reader}. Reader has obtained wavefunctions using
fitting parameters by comparing calculated and experimental
energies, whereas, Pan and Beck have used relativistic CI method
for the available transitions. Our calculated values of oscillator
strength for the 4d $^{2}D_{3/2}$  $\rightarrow$ 4f $^{2}F_{5/2}$
transition at the DF level given in the table, agrees well with
similar calculations by Zilitis \cite{zilitis}, 1.023.

\begin{table}
\caption{Oscillator strengths for E1 transitions in length form
and their comparison with earlier results.}
\begin{tabular}{lclrrrr}
\hline
Term &  &        &  \multicolumn {2}{c} {Present calculations}   & \multicolumn {2}{c} {Other calculations}   \\
Upper & & Lower  & DF  & CC & \cite{reader} & \cite{pan} \\
\hline \hline
4d $^{2}D_{3/2}$ & $\rightarrow$ & 4f $^{2}F_{5/2}$ & 1.0099   & 0.3967 & 0.3226 & 0.2896  \\
4d $^{2}D_{5/2}$ & $\rightarrow$ & 4f $^{2}F_{5/2}$ & 0.0484 & 0.0246 & 0.0153 & 0.0139  \\
\hline
\end{tabular}
\label{tab:results1}
\end{table}

Weighted oscillator strengths correspond to E1 transitions are
presented in Table IV. Here we have used the length gauge values
of E1 transition amplitudes and our calculated wavelengths. All
these transitions, fall in ultraviolet and visible regions, are
useful for astrophysical observations and may be for laboratory
researches. According to Cowan \cite{cowan}, if the initial states
are dominated by $^2D$ and final states are dominated by $^2P$,
the oscillator strength ratio of $^2D_{5/2} \rightarrow
^2P_{3/2}$, $^2D_{3/2} \rightarrow ^2P_{1/2}$ and $^2D_{3/2}
\rightarrow ^2P_{3/2}$ transitions in a given multiplet are 6:5:1,
which we find the same for 4d $^2D$ $\rightarrow$ 5p $^2P$
transitions.

\begin{table}
\caption{Transition wavelengths  (in nm) and weighted oscillator
strengths (gf) corresponding to electric dipole (E1) transitions
of Mo VI.}
\begin{tabular}{lcrrrr}
\hline
Term &     &   &       &        &        \\
Upper        &     & Lower              &  $\lambda_{exp.}$ & gf
\\ \hline \hline
5p $^{2}P_{1/2}$ & $\rightarrow$ & 4d $^{2}D_{3/2}$ &  54.82 & 0.5354 \\
5P $^{2}P_{1/2}$ & $\rightarrow$ & 5s $^{2}S_{1/2}$ & 159.54 & 0.5920\\
5p $^{2}P_{3/2}$ & $\rightarrow$ & 4d $^{2}D_{3/2}$ &  53.38 & 0.1042 \\
5p $^{2}P_{3/2}$ & $\rightarrow$ & 4d $^{2}D_{5/2}$ &  54.12 & 0.9676 \\
5p $^{2}P_{3/2}$ & $\rightarrow$ & 5s $^{2}S_{1/2}$ & 147.91 & 1.2844 \\
4f $^{2}F_{5/2}$ & $\rightarrow$ & 4d $^{2}D_{3/2}$ &  37.44 & 1.5868 \\
4f $^{2}F_{5/2}$ & $\rightarrow$ & 4d $^{2}D_{5/2}$ &  37.81 & 0.1478 \\
4f $^{2}F_{7/2}$ & $\rightarrow$ & 4d $^{2}D_{5/2}$ &  37.75 & 5.7164 \\
5d $^{2}D_{3/2}$ & $\rightarrow$ & 5p $^{2}P_{1/2}$ &  99.58 & 2.3133 \\
5d $^{2}D_{3/2}$ & $\rightarrow$ & 5p $^{2}P_{3/2}$ & 104.71 & 0.4613 \\
5d $^{2}D_{3/2}$ & $\rightarrow$ & 4f $^{2}F_{5/2}$ & 633.77 & 0.3446 \\
5d $^{2}D_{5/2}$ & $\rightarrow$ & 5p $^{2}P_{3/2}$ & 103.86 & 4.5998 \\
5d $^{2}D_{5/2}$ & $\rightarrow$ & 4f $^{2}F_{5/2}$ & 603.72 & 0.0260 \\
5d $^{2}D_{5/2}$ & $\rightarrow$ & 4f $^{2}F_{7/2}$ & 619.03 & 0.5054 \\
6s $^{2}S_{1/2}$ & $\rightarrow$ & 5p $^{2}P_{1/2}$ & 76.10 & 0.3881\\
6s $^{2}S_{1/2}$ & $\rightarrow$ & 5p $^{2}P_{3/2}$ & 79.06 & 0.8468 \\
6p $^{2}P_{1/2}$ & $\rightarrow$ & 4d $^{2}D_{3/2}$ &  29.36 & 0.0409 \\
6p $^{2}P_{1/2}$ & $\rightarrow$ & 5s $^{2}S_{1/2}$ &  45.28 & 0.0415 \\
6p $^{2}P_{1/2}$ & $\rightarrow$ & 5d $^{2}D_{3/2}$ &  173.17 & 1.1894 \\
6p $^{2}P_{1/2}$ & $\rightarrow$ & 6s $^{2}S_{1/2}$ &  373.63 & 0.9882 \\
6p $^{2}P_{3/2}$ & $\rightarrow$ & 4d $^{2}D_{3/2}$ & 29.19 & 0.0092\\
6p $^{2}P_{3/2}$ & $\rightarrow$ & 4d $^{2}D_{5/2}$ &  29.41 & 0.0841 \\
6p $^{2}P_{3/2}$ & $\rightarrow$ & 5s $^{2}S_{1/2}$ &  44.87 & 0.0559 \\
6p $^{2}P_{3/2}$ & $\rightarrow$ & 5d $^{2}D_{3/2}$ & 167.40 & 0.2311 \\
6p $^{2}P_{3/2}$ & $\rightarrow$ & 5d $^{2}D_{5/2}$ &  169.63 &  2.0902 \\
6p $^{2}P_{3/2}$ & $\rightarrow$ & 6s $^{2}S_{1/2}$ & 347.75 & 2.1442 \\
\hline
\end{tabular}
\label{tab:results1}
\end{table}


In Table V, we present M1 and E2 transition probabilities and
their corresponding wavelengths. However most of the transitions
come in ultraviolet region, there are few transitions fall in
infrared region. Though these transitions produce weak
lines but they are important parameters in
astrophysical studies. As expected, transition probability for E2 transitions
come greater in value than that of M1 transitions except for the
transitions fall in infrared region.

\begin{table}
\caption{Transition wavelengths  (in nm) and transition
probabilities corresponding to electric quadrupole (E2) and
magnetic dipole transitions (M1) (in sec$^{-1}$) of Mo VI.}
\begin{tabular}{lcrrrr}
\hline
Term &     &   &       &        &        \\
Upper & & Lower  &  $\lambda_{exp.}$ & A$_{E2}$ & A$_{M1}$ \\
\hline \hline
4d $^{2}D_{5/2}$ & $\rightarrow$ & 4d $^{2}D_{3/2}$ & 3870.71 &  3.9372$\times$10$^{-6}$ & 2.006$\times$10$^{-1}$ \\
5s $^{2}S_{1/2}$ & $\rightarrow$ & 4d $^{2}D_{3/2}$ &   83.52 & 6.9919$\times$10$^{3}$ & 2.8677$\times$10$^{-7}$ \\
5s $^{2}S_{1/2}$ & $\rightarrow$ & 4d $^{2}D_{5/2}$ &   85.36 & 9.3441$\times$10$^{3}$ &  \\
5p $^{2}P_{3/2}$ & $\rightarrow$ & 5p $^{2}P_{1/2}$ & 2029.74 & 4.2402$\times$10$^{-3}$ & 1.1401 \\
4f $^{2}F_{5/2}$ & $\rightarrow$ & 5p $^{2}P_{1/2}$ &  118.14 & 4.5388$\times$10$^{3}$ &  \\
4f $^{2}F_{5/2}$ & $\rightarrow$ & 5p $^{2}P_{3/2}$ &  125.44 & 9.8580$\times$10$^{2}$ & 1.6191$\times$10$^{-5}$ \\
4f $^{2}F_{7/2}$ & $\rightarrow$ & 5p $^{2}P_{3/2}$ &  124.80 & 4.2258$\times$10$^{3}$ &  \\
4f $^{2}F_{7/2}$ & $\rightarrow$ & 4f $^{2}F_{5/2}$ &24412.87 & 7.5555$\times$10$^{-11}$ & 1.3193$\times$10$^{-4}$ \\
5d $^{2}D_{3/2}$ & $\rightarrow$ & 4d $^{2}D_{3/2}$ &   35.35 & 1.1221$\times$10$^{5}$ & 6.4795$\times$10$^{-5}$ \\
5d $^{2}D_{3/2}$ & $\rightarrow$ & 4d $^{2}D_{5/2}$ &   35.68 & 4.8435$\times$10$^{4}$ & 2.8382$\times$10$^{1}$ \\
5d $^{2}D_{3/2}$ & $\rightarrow$ & 5s $^{2}S_{1/2}$ &   61.31 & 1.2090$\times$10$^{5}$ & 5.7627$\times$10$^{-5}$ \\
5d $^{2}D_{5/2}$ & $\rightarrow$ & 4d $^{2}D_{3/2}$ &   35.25 & 3.1111$\times$10$^{4}$ & 2.5858 \\
5d $^{2}D_{5/2}$ & $\rightarrow$ & 4d $^{2}D_{5/2}$ &   35.58 & 1.2746$\times$10$^{5}$ & 1.6034$\times$10$^{1}$\\
5d $^{2}D_{5/2}$ & $\rightarrow$ & 5s $^{2}S_{1/2}$ &   61.01 & 1.2112$\times$10$^{5}$ &  \\
5d $^{2}D_{5/2}$ & $\rightarrow$ & 5d $^{2}D_{3/2}$ & 12733.17 & 2.9790$\times$10$^{-7}$ & 5.3860$\times$10$^{-3}$ \\
6s $^{2}S_{1/2}$ & $\rightarrow$ & 4d $^{2}D_{3/2}$ &   31.86 & 1.7297$\times$10$^{2}$ & 1.5970$\times$10$^{-3}$\\
6s $^{2}S_{1/2}$ & $\rightarrow$ & 4d $^{2}D_{5/2}$ &   32.13  & 8.9836$\times$10$^{2}$ &  \\
6s $^{2}S_{1/2}$ & $\rightarrow$ & 5s $^{2}S_{1/2}$ &   51.52 &  & 2.9841 \\
6s $^{2}S_{1/2}$ & $\rightarrow$ & 5d $^{2}D_{3/2}$ &   322.77 & 2.6456$\times$10$^{2}$ & 2.6717$\times$10$^{-6}$ \\
6s $^{2}S_{1/2}$ & $\rightarrow$ & 5d $^{2}D_{5/2}$ &   331.17 & 3.5638$\times$10$^{2}$ &  \\
6p $^{2}P_{1/2}$ & $\rightarrow$ & 5p $^{2}P_{1/2}$ & 63.22  &  & 8.4454$\times$10$^{-2}$ \\
6p $^{2}P_{1/2}$ & $\rightarrow$ & 5p $^{2}P_{3/2}$ &  65.25 & 7.4911$\times$10$^{4}$ & 4.0073$\times$10$^{2}$ \\
6p $^{2}P_{1/2}$ & $\rightarrow$ & 4f $^{2}F_{5/2}$ &  136.01 & 2.2749$\times$10$^{3}$ &  \\
6p $^{2}P_{3/2}$ & $\rightarrow$ & 5p $^{2}P_{1/2}$ &  62.44 & 3.6089$\times$10$^{4}$ & 1.3776$\times$10$^{2}$ \\
6p $^{2}P_{3/2}$ & $\rightarrow$ & 5p $^{2}P_{3/2}$ & 64.42 & 3.5807$\times$10$^{4}$ & 1.6161 \\
6p $^{2}P_{3/2}$ & $\rightarrow$ & 4f $^{2}F_{5/2}$ &   132.42 & 3.5357$\times$10$^{2}$ & 3.1526$\times$10$^{-7}$  \\
6p $^{2}P_{3/2}$ & $\rightarrow$ & 4f $^{2}F_{7/2}$ &   131.34 & 2.0505$\times$10$^{3}$ &  \\
6p $^{2}P_{3/2}$ & $\rightarrow$ & 6p $^{2}P_{1/2}$ & 5020.93 &  9.9619$\times$10$^{-4}$ & 1.1185$\times$10$^{-1}$ \\
\hline
\end{tabular}
\label{tab:results1}
\end{table}

In Table VI, the contributions due to the different correlation
terms like core-correlation, pair-correlation, core-polarization
and important two-body contributions are estimated for few
transitions to highlight the effect of correlations. Significant
correlation contributions from the higher order core-polarization,
like $S^{\dag (0,1)}_{2f}\bar{O}S_{2i}^{(0,1)}$, are noticeable
compared to the lowest order contributions. Also, contributions
from two-body correlations are almost comparable for most of the
cases. For the M1 transition, 4d $^{2}D_{5/2}$ $\rightarrow$ 4d
$^{2}D_{3/2}$, there are cancelation observed among different
correlation effects. The contribution comes from core-polarization
term is more compared to other term. Even in the case of 5s
$^{2}S_{1/2}$ $\rightarrow$ 4d $^{2}D_{3/2}$ M1 transition, the
Dirac-Fock contribution is almost canceled by the lowest order of
core-polarization, which makes the core-corelation effect more
dominant to the total value of transition matrix element.

\begin{table}
\scriptsize
\caption{Explicit contributions from the CCSD(T) calculations to the absolute magnitude of transition amplitudes.}
\begin{tabular}{lrrrrrrr}
\hline & Dirac-Fock & Core-corr. & Pair-corr. & Core-polar.  & Core-polar.  & Two-body  & Total     \\
       &            &            &            &    (lowest)          & (higher)             & contr.   & \\
\hline \hline
5p(1/2) $\rightarrow$ 4d(3/2) (E1) & -1.0898 &  -2.1536E-3 & 2.4778E-2  & 8.1783E-2 & -8.4547E-3 & -8.5320E-4 & -0.9851 \\
5p(3/2) $\rightarrow$ 4d(3/2) (E1) & -0.4704  & -8.5909E-4 & 1.1185E-2  & 3.1035E-2 & -3.4575E-3 & -3.6146E-4 & -0.4288 \\
4d(5/2) $ \rightarrow$ 4d(3/2) (E2) & 1.4451  & -2.5900E-2 & -3.1118E-2 & -1.3272E-1 & 3.6817E-3 & -1.1045E-3 & 1.2432 \\
5s(1/2) $\rightarrow$ 4d(3/2) (E2) & 2.4193  & 2.8071E-3 & -6.8921E-2 & -3.1504E-2 & 2.1471E-2 & -2.6223E-4 & 2.3176 \\
4d(5/2) $\rightarrow$ 4d(3/2) (M1) & 1.5488  & -1.3535E-2 &
-1.4813E-4 & 1.5097E-4 & 1.2398E-2 &
-1.8252E-3 & 1.5284 \\
5s(1/2) $\rightarrow$ 4d(3/2) (M1) & -2.5086E-5  & 9.4303E-6 & -3.1876E-7 & 2.4685E-5 & -4.8217E-6 & 6.6800E-9 & 3.5805E-6 \\
\hline
\end{tabular}
\label{tab:results1}
\end{table}

Table VII summarizes
the calculated lifetime of the low-lying excited states.
Recent calculations  of the lifetimes
of 5p $^2P_{1/2}$ and 5p $^2P_{3/2}$ states by Zilitis \cite{zilitis} are
also compared here. Here we can see that the lifetime of the 4d
$^2D_{5/2}$ state comes in the order of second, which suggest that Mo VI can be used for
uniform doping in thin film. The lifetime of the 5s $^2S_{1/2}$ state
is found to be of the order of microsecond due to only forbidden transition contributions. Lifetime of the $5d$
states are larger than the $4f$ states due to strong contributions from the allowed dipole
transitions 4f $^2F$ $\rightarrow$ 4d $^2D$.

\begin{table}
\caption{Radiative lifetimes (in sec.) for different low-lying
states of Mo VI.}
\begin{tabular}{lrr}
\hline Term & Present calculations & Other calculations$^a$ \\
\hline \hline
4d $^{2}D_{5/2}$ &  4.9854 & \\
5s $^{2}S_{1/2}$ & 6.1213$\times$10$^{-5}$ & \\
5p $^{2}P_{1/2}$ & 1.4968$\times$10$^{-10}$ & 1.300$\times$10$^{-10}$\\
5p $^{2}P_{3/2}$ & 1.4151$\times$10$^{-10}$ & 1.260$\times$10$^{-10}$\\
4f $^{2}F_{5/2}$ & 7.1628$\times$10$^{-11}$ & \\
4f $^{2}F_{7/2}$ & 2.9462$\times$10$^{-11}$ & \\
5d $^{2}D_{3/2}$ & 2.1575$\times$10$^{-10}$ & \\
5d $^{2}D_{5/2}$ & 2.0922$\times$10$^{-10}$ & \\
\hline
\end{tabular}
\label{tab:results1}

$^a$ Ref. \cite{zilitis}. \\
\end{table}

\section{Conclusion}
Forbidden transition probabilities among the low-lying states of Mo VI
relevant for astro- and plasma physics are calculated using highly
correlated relativistic coupled-cluster method for the first time in
literature to the best
of our knowledge. The lifetime of the 4d $^2D_{5/2}$ state is
found to be around 5 second, which will be useful in many physical processes.
Contributions of different correlation
terms are discussed and found strong effect from higher order
core-polarization. In the near future, present work will motivate
experimentalists to verify our results due to its importance in many areas in physics.

\section{Acknowledgment}
We are grateful to Prof B P Das and Dr Rajat K Chaudhuri , Indian
Institute of Astrophysics, Bangalore  for providing the CC code. One
of us (Narendra) would like to recognize the support of Council of
Scientific and Industrial Research (CSIR), India.

\end{document}